\input{aipcheck}

\documentclass[
    ,final            
  ]
  {aipproc}

\layoutstyle{6x9}

\begin{document}

\title{Uncertainties and Systematic Effects on the estimate of stellar masses in high z galaxies.}


\author{S. Salimbeni}{
  address={Department of Astronomy-UMASS-710 North Pleasant St., Amherst, MA, 01003}
  ,altaddress={INAF-OAR, Via frascati 33, Monte Porzio Catone, 00040, Italy} 
}

\author{A. Fontana}{
  address={INAF-OAR, Via frascati 33, Monte Porzio Catone, 00040, Italy}
}
\author{E.Giallongo}{
  address={INAF-OAR, Via frascati 33, Monte Porzio Catone, 00040, Italy}
}
\author{A.Grazian}{
  address={INAF-OAR, Via frascati 33, Monte Porzio Catone, 00040, Italy}
}
\author{N.Menci}{
  address={INAF-OAR, Via frascati 33, Monte Porzio Catone, 00040, Italy}
}
\author{L.Pentericci}{
  address={INAF-OAR, Via frascati 33, Monte Porzio Catone, 00040, Italy}
}
\author{P.Santini}{
  address={INAF-OAR, Via frascati 33, Monte Porzio Catone, 00040, Italy}
}

\begin{abstract}
  We discuss the uncertainties and the systematic effects that exist
  in the estimates of the stellar masses of high redshift galaxies,
  using broad band photometry, and how they affect the deduced galaxy
  stellar mass function. We use at this purpose the latest version of
  the GOODS-MUSIC catalog. In particular, we discuss the impact of
  different synthetic models, of the assumed initial mass function and
  of the selection band.  Using Charlot \& Bruzual 2007 and Maraston
  2005 models we find masses lower than those obtained from Bruzual \&
  Charlot 2003 models. In addition, we find a slight trend as a
  function of the mass itself comparing these two mass determinations
  with that from Bruzual \& Charlot 2003 models. As consequence, the
  derived galaxy stellar mass functions show diverse shapes, and their slope depends on the assumed models. Despite these differences, the overall results and scenario remains unchanged. The masses obtained with the assumption
  of the Chabrier initial mass function are in average 0.24 dex lower
  than those from the Salpeter assumption, at all redshifts, causing a
  shift of galaxy stellar mass function of the same amount.  Finally,
  using a 4.5 $\mu m$-selected sample instead of a Ks-selected one, we
  add a new population of highly absorbed, dusty galaxies at $z\simeq
  2-3$ of relatively low masses, yielding stronger constraints on the
  slope of the galaxy stellar mass function at lower
  masses.  \end{abstract}

\maketitle


\section{Introduction}

Galaxy stellar mass (GSM) and galaxy stellar mass
function (GSMF) has been the main topics of numerous works in recent
years \citep[e.g.,][]{fontana06,pozzetti2007,
  perez2008,marchesini2008}.  So far a downsizing scenario has
emerged: most massive galaxies ($M>10^{11} M_\odot$) had faster
evolution at high redshift, with about 50\% of their mass already
assembled at $z\sim 1$; instead, the mass assembly in less massive
galaxies continues up to low redshift \citep[e.g.][]{fontana06}. These
analysis use multicolor catalogs, and the GSM estimates are inferred
from the comparison between spectral energy distribution (SED) and
libraries of synthetic spectra. Assumptions and systematics affect
these estimates and the derived GSMF. Among the others, systematics
originate from the assumed ingredients in the building of the library
of synthetic spectra (e.g. stellar population models, initial mass
function (IMF), SFR histories, reddening).  Furthermore, as the $M/L$
ratio depends from the spectral type, the selection band could affect
the GSMF determination (for a detailed description of this problem see
the related section and \citep{fontana2004,fontana06}).
	
To discuss the impact of these uncertainties, we use the latest
version of the GOODS--MUSIC multi-wavelength catalog, covering 14
bands from U to $8 \mu m$ band (Santini et al. submitted). We show how
the assumptions on the stellar population models and IMF affect the
results. Using a 4.5$\mu m$-selected sample
($4.5\mu m \sim 23.1$ ), we also show a comparison between GSMFs
obtained using different selection bands ($Ks$ and 4.5$\mu m$).

\section{Different recipes for stellar mass estimates}

\begin{figure}
  \includegraphics[width=.8\textwidth]{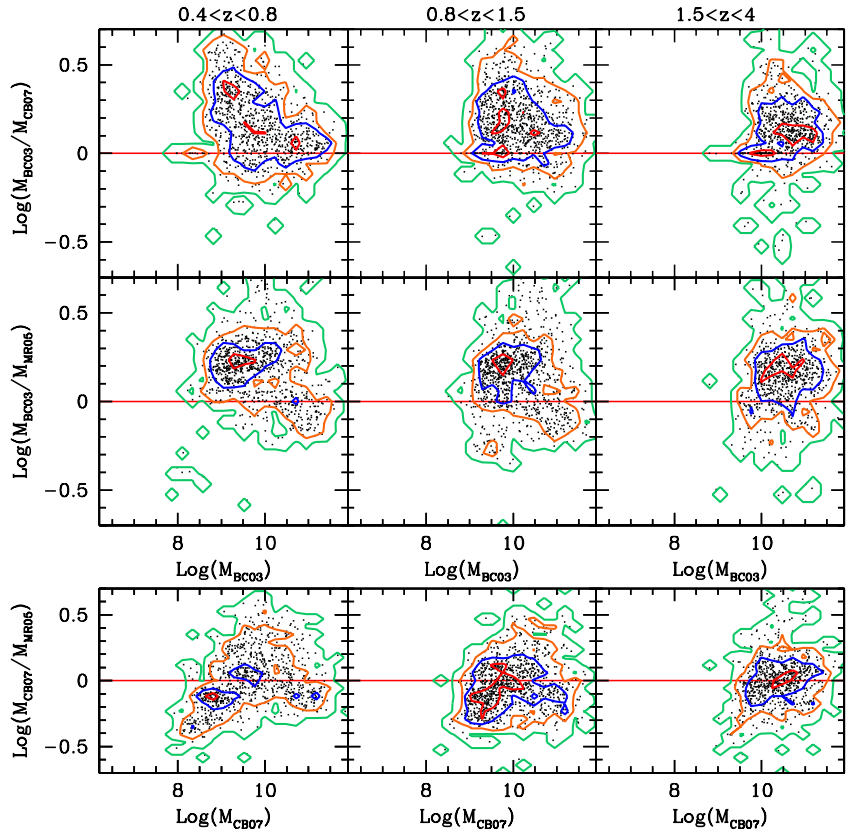}
  \caption{Comparison of the masses obtained using BC03, MR05 and CB07 models. Continuos line are the isodensity contours indicating  0.01,0.1,0.3 and 0.7 of the maximal density of galaxies.}\label{fig:1}
\end{figure}

\begin{figure}
    \includegraphics[height=.43\textheight]{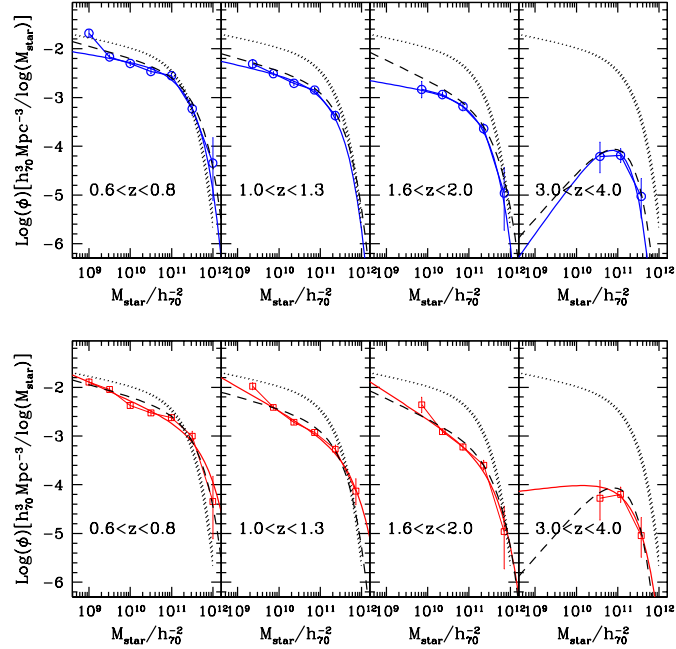}
   \caption{GSMFs obtained in different redshift intervals for the MR05 (red continuous lines and empty square points) and CB07 (blue continuous lines and empty circle points)  models. The dashed lines are GSMFS obtained with the BC03 models. The shaded area indicates the local GSMF by \cite{cole2001}.}\label{fig:2}
 \end{figure}

The procedure adopted to obtain the GSMs is described in detail in
\cite{fontana2004,fontana06}. In this section we show how the inferred
masses change adopting the same initial conditions (Salpeter IMF
\citep{salpeter1955}, metallicity and age grids, etc...) but different
simple stellar population tracks.  In Fig. \ref{fig:1}, we show the comparison between masses obtained using
the Bruzual \& Charlot 2003, Maraston 2005
\citep{bruzual2003,maraston2005} and Charlot \& Bruzual 2007
\cite{bruzual2007a, bruzual2007b} models (elsewhere BC03, MR05 and
CB07).  The continuos lines of different colors indicate isodensity
contours (0.01,0.1,0.3,0.7 of the maximal density).  As expected, the
masses obtained with MR05 and CB07 are on average lower than those
obtained with BC03 \citep{bruzual2007a,bruzual2007b,maraston2006}.
However, as it is shown in the redshift bin 0.4-0.8, the ratio between
CB07 and BC03 shows a dependence with the mass itself. Masses obtained
from BC03 models are higher than the CB07 ones of 0.3 dex at $M<9.5
M_\odot$, and of 0.05 dex at $M>10^{11} M_\odot$. Instead, the MR05
model gives masses in average lower than those from BC03 of a constant
amount ($\sim 0.2 $ dex) with the exception of a small group of
objects at $M>10^{11} M_\odot$.  In the higher redshift bins the mass
range covered is smaller, and this makes the effect described above
less evident. These systematic in the mass estimates cause different
shapes in GSMFs and affect their slopes, as shown in Fig. \ref{fig:2}. The result is a shallower GSMF for the CB07 models and a steeper one for MR05 with respect to the determination obtained using the BC03 models. However, the general trend of fast evolution at high redshift for the most massive galaxies ($M>10^{11} M_\odot$)  is observed, as
shown in Fig. \ref{fig:2}.
 
\begin{figure}
  \includegraphics[height=.43\textheight]{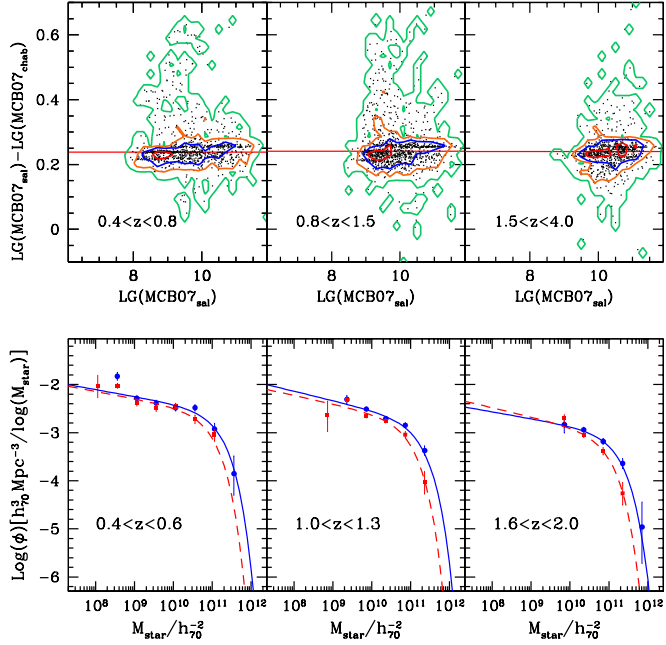}
  \caption{Top: comparison between masses obtained with a Chabrier IMF and with a Salpeter one, the curves are the same of Fig.  \ref{fig:1} .  Bottom: comparison of the GSMF obtained with a Chabrir IMF (blue continuos curve and points) and with a Salpeter one (red dashed line and  points).}\label{fig:3}
\end{figure}

We also tested how the GSMF changes under different assumptions of the
IMF. In particular we adopt the Chabrier \citep{chabrier2003} and the
Salpeter IMFs, the former being steeper at low stellar masses than the
latter. This different shape causes a systematic in the GSMs as shown
in Fig. \ref{fig:3}. Here, in the upper panels, we
compare masses obtained with our standard recipes and CB07 libraries,
but with the two different choice of IMF. The masses obtained using a
Chabrier IMF are in average 0.24 dex lower than those obtained
adopting a Salpeter IMF, and this effect is totally independent from
the range of redshift and mass probed. As a consequence the GSMFs have the same
shape, but the one obtained with a Chabrier IMF is shifted at lower
masses of the same amount above mentioned, as is shown in the bottom
panel of Fig. \ref{fig:3}.

\section{Mass function with a 4.5 $\mu m$ -selected sample}

\begin{figure}
  \includegraphics[height=.40\textheight]{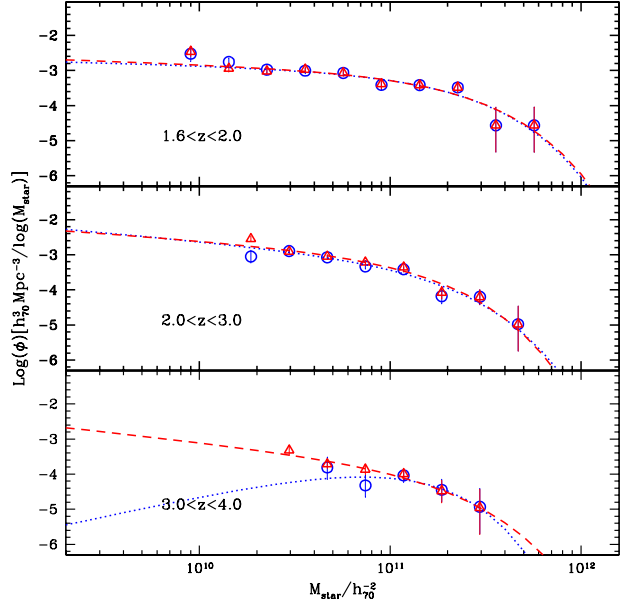}
  \caption{GSMFs obtained with different  magnitude-selected samples. Red dashed line from a 4.5$\mu m$-selected sample, and the blue one from a $Ks$-selected sample.}\label{fig:4}
\end{figure}

The effects described in the previous section depend on different
assumptions of the library ingredients. Here, we focus on the selection
effects of our sample.  The use of a magnitude limited sample does not
translate into a well defined, sharp limit in stellar mass, as the M/L
ratio depends from the galaxy spectral type. In the case of a
$Ks$-magnitude limited catalog, the sample can be complete for early
type galaxies and moderately obscured star-forming ones, while it
could miss very dusty star-forming galaxies, as widely discussed in
\citep{fontana06}. As result, the $Ks$-selected GOODS--MUSIC
sample could be affected by this kind of incompleteness over redshift
$\sim 2$ \citep[see Fig. 3 in][]{fontana06}.  This problem could be
alleviated by a 4.5 $\mu m$-selection, which probes longer wavelengths
less affected by dust extinction at these redshifts. Adding the
objects selected in the $4.5\mu$m band, but fainter in $K$, we find a
population of highly absorbed, dusty galaxies at $z\simeq 2-3$ of
relatively low masses. The effect on GSMF is shown in Fig. \ref{fig:4}, showing the comparison of the GSMF obtained from
$Ks$ (blue dotted-lines and empty circle points) and 4.5 $\mu m$ (red
dashed-lines and empty triangular points) sample. We find that this new
selection does not increase the GSMF at higher masses and that very
dusty galaxies do not give a relevant contribution to the amount of
stellar mass at all redshift considered here. At lower masses, in the
redshift bins between 1.6 and 3, it improves the statistics, and
although it does not change the results, it helps given a more solid
constraint to the determination of $\alpha$ parameter of the Schechter
function.  This contribution becomes essential in the determination of
the $\alpha$ parameter in the higher redshift bin.

\hyphenation{Post-Script Sprin-ger}


\end{document}